\documentclass{egpubl}
\usepackage{eurovis2024}

\SpecialIssuePaper         

\CGFccby

\usepackage[T1]{fontenc}
\usepackage{dfadobe}

\usepackage{cite}  
\BibtexOrBiblatex
\electronicVersion
\PrintedOrElectronic
\ifpdf \usepackage[pdftex]{graphicx} \pdfcompresslevel=9
\else \usepackage[dvips]{graphicx} \fi

\usepackage{egweblnk}
\usepackage{mathtools}
\usepackage{enumitem}
\usepackage[capitalise,noabbrev]{cleveref}

\newcommand{\AB}{\mathbf{A}}

\newcommand{\PP}{\mathbf{P}}
\newcommand{\pp}{\mathbf{p}}

\DeclareMathOperator*{\argmin}{arg\,min}

\title[Instantaneous Visual Analysis of Blood Flow in Stenoses Using Morphological Similarity]%
      {Instantaneous Visual Analysis of Blood Flow in Stenoses Using Morphological Similarity}

\author[P. Eulzer et al.]{
{\parbox{\textwidth}{\centering P. Eulzer$^{1}$, K. Richter$^{2}$, A. Hundertmark$^{2}$, R. Wickenh{\"o}fer$^{3}$, C.\,M. Klingner$^{4}$ and K. Lawonn$^{1}$}}
\\
{\parbox{\textwidth}{\centering $^1$University of Jena, Faculty of Mathematics and Computer Science, Germany\\
         $^2$ RPTU Kaiserslautern-Landau, Institute of Mathematics, Germany\\
         $^3$ Herz-Jesu Hospital Dernbach, Clinic for Radiology, Germany\\
         $^4$ University Hospital Jena, Clinic for Neurology, Germany
       } 
}}

\begin{document}

\teaser{
  \includegraphics[width=0.98\linewidth]{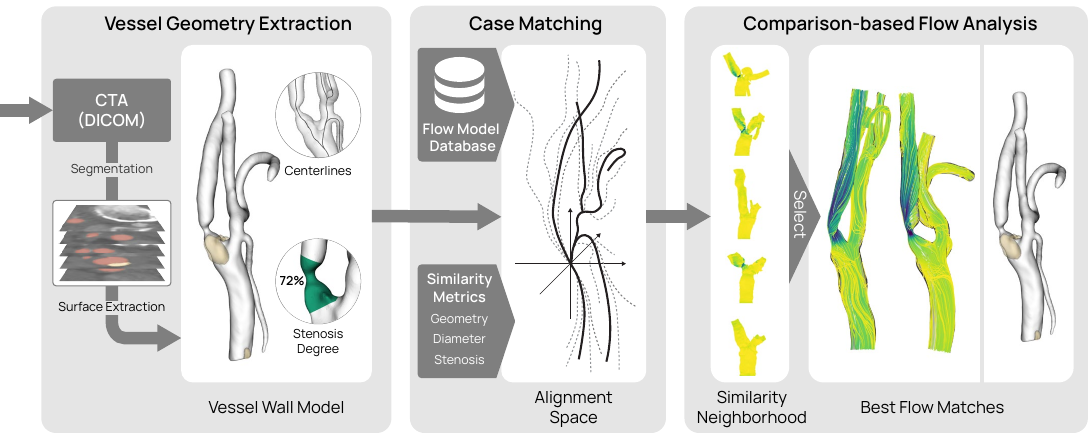}
  \centering
  \caption{Overview of the data processing and visualization steps. The vessel geometry and features are extracted from an imported DICOM series. Using similarity metrics, the new model is matched against a flow model database. Then, we use comparative visualization techniques to enable the exploration of flow parameters in the best matching vessels.}
  \label{fig:teaser}
}

\maketitle
\begin{abstract}
  The emergence of computational fluid dynamics (CFD) enabled the simulation of intricate transport processes, including flow in physiological structures, such as blood vessels.
  While these so-called hemodynamic simulations offer groundbreaking opportunities to solve problems at the clinical forefront, a successful translation of CFD to clinical decision-making is challenging.
  Hemodynamic simulations are intrinsically complex, time-consuming, and resource-intensive, which conflicts with the time-sensitive nature of clinical workflows and the fact that hospitals usually do not have the necessary resources or infrastructure to support CFD simulations.
  To address these transfer challenges, we propose a novel visualization system which enables instant flow exploration without performing on-site simulation.
  To gain insights into the viability of the approach, we focus on hemodynamic simulations of the carotid bifurcation, which is a highly relevant arterial subtree in stroke diagnostics and prevention.
  %
  %
  We created an initial database of 120 high-resolution carotid bifurcation flow models and developed a set of similarity metrics used to place a new carotid surface model into a neighborhood of simulated cases with the highest geometric similarity.
  %
  The neighborhood can be immediately explored and the flow fields analyzed.
  %
  %
  We found that if the artery models are similar enough in the regions of interest, a new simulation leads to coinciding results, allowing the user to circumvent individual flow simulations.
  We conclude that similarity-based visual analysis is a promising approach toward the usability of CFD in medical practice.


\end{abstract}

\section{Introduction}
The effective treatment and prevention of stroke is of high societal importance~\cite{gorelick2019global}.
Stroke is the second leading cause of death globally~\cite{burden_of_stroke} and post-stroke care is linked with considerable economic expenses~\cite{rajsic2019economic}.
A significant risk factor for stroke is carotid stenosis, which describes a narrowing of the carotid arteries.
The carotids are two arteries located on each side of the neck and are the main source of blood flow to the brain.
When these arteries become narrowed or blocked, it can lead to a reduction in blood flow to the brain, which increases the risk of stroke.
Carotid stenosis is commonly diagnosed using computed tomography angiography (CTA) and Doppler ultrasonography.
CTA provides a static volume image of the head and neck region with highlighted arteries.
Doppler ultrasonography is used to get a dynamic view of the blood flow and to measure the flow velocity.
The blood flow velocity increases with a smaller vessel diameter, and as such it is used as a primary clinical marker for the severeness of the stenosis.
A disadvantage of these methods is that the data from the CTA and sonography need to be mentally combined.
Two separate procedures, involving different experts and equipment need to be performed.
Doppler sonography, in particular, requires a highly trained medical professional to be carried out correctly.
The necessary personnel and devices are not available in many smaller clinics, which means measuring flow can be inaccessible.

In the last decades, a large body of research has been dedicated to simulating arterial blood flow~\cite{bluestein2017utilizing, caballero2013review, LOPES2020110019, SZAJER201862}.
These simulations are numerical predictions commonly using computational fluid dynamics (CFD) to derive flow properties based on the vessel wall geometry.
This geometry can be extracted from static imaging like CTA, theoretically omitting the need for additional diagnostic procedures.
Hemodynamic simulation can also yield a much higher resolution than sonography and it can be applied to regions where ultrasound signals are occluded.
It can further be used to derive many additional parameters beyond flow velocity, where some have been shown to be relevant biomarkers for predicting the progression of atherosclerotic diseases~\cite{hartman2021definition, nisco2020impact}.

While hemodynamic simulations have the potential to significantly impact patient care and treatment outcomes, they are difficult to integrate into clinical workflows, due to the often time-consuming, resource-intensive, and complex nature of the simulations.
The simulations used in this work, for instance, required an average computation time of 4 hours and 50 minutes per artery, excluding pre- and post-processing steps like geometry extraction.
Hemodynamic simulations are often developed by specialized research groups and require significant expertise, as well as specific software and hardware to perform.
In this work, we propose a method to circumvent performing hemodynamic simulation for each new case individually.
An overview of our approach is shown in \cref{fig:teaser}.
We created a database of high-resolution flow models for a wide range of carotid bifurcation geometries.
Then, we developed a framework which, given a new carotid wall geometry, finds the candidates with highest similarity from the simulation database.
We investigated how similarity between vascular models can be established and whether similarity-based analysis is a viable approach to circumvent complex individual flow simulations.
We enabled exploring multiple candidates and visually comparing their geometries, simulated flow, and wall stress.
%
%
In summary, our contributions are:
\begin{itemize}
    \item Creating an open-source database of 120 high-resolution carotid bifurcation flow models extracted from individual patients.
    \item Developing difference metrics to create an alignment space and derive model similarity.
    \item Implementing and testing an interactive framework that, given a new vessel wall model, can be used to instantly find and visualize the most fitting flow simulations.
\end{itemize}
The developed software is open source and implemented as an extension module of the CarotidAnalyzer framework~\cite{CarotidAnalyzer}.
The carotid flow database is available online~\cite{eulzer2024dataset}.
\section{Medical Background}\label{sec:medical_background}
\begin{figure}[t]
\centering
\includegraphics[width=0.8\linewidth]{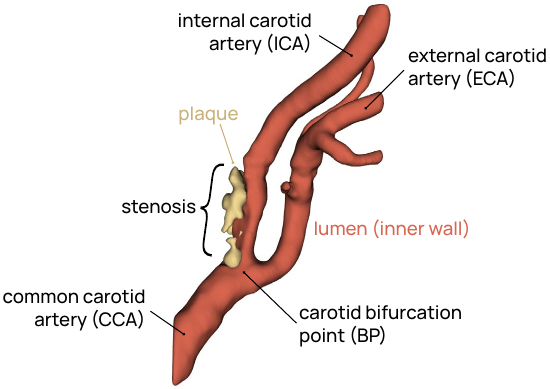}
\caption{Carotid bifurcation anatomy.}
\label{fig:anatomy}
\end{figure}

\noindent The right and left common carotid arteries (CCA) originate from the aorta.
At the carotid bifurcation, they divide into two branches, the internal carotid artery (ICA) and external carotid artery (ECA), see \cref{fig:anatomy}.
The ICA is responsible for supplying oxygen-rich blood to the brain and any blockage or narrowing, i.e., stenosis, can result in a stroke.
Carotid stenosis is usually caused by the buildup of fatty deposits called plaques in the inner lining of the artery.
As these plaques grow, they can constrict the artery and reduce blood flow.
If a piece of the plaque breaks off, it can form a clot that often blocks a smaller artery upstream, also leading to a stroke.
One way to prevent this is to identify developing stenoses early and perform stent insertion or surgical removal of plaque to avoid total closure and the resulting blockage of smaller vessels.
However, carotid surgery has its own risks that must be carefully considered against the likelihood of a stroke~\cite{halm2009risk}.
Therefore, selecting the right moment for surgery and the best treatment strategy is critical, which requires careful evaluation of each case and stenotic region.
Current clinical guidelines base the treatment recommendations largely on the stenosis degree~\cite{NASCET}, which can be derived by measuring diameters of the vessel lumen (the inner wall) on an angiographic image like a CTA.
However, using the stenosis degree as a threshold value alone, does not include other factors, such as the individual shape of the stenosis, the plaque type and distribution, or the remaining blood flow rate.
Since parameters like the flow velocity are not available from static CTA imaging but require an additional Doppler ultrasonography, an approximation of the hemodynamics from CTA could accelerate the evaluation of these cases.
Furthermore, studies based on hemodynamic simulation show that carotid bifurcation plaque formation is influenced by vascular wall shear stress (WSS), dynamic pressure, strain rate, and the total pressure gradient~\cite{li2019hemodynamic}, all of which cannot be measured from CTA or Doppler sonography.
As hemodynamic simulations are not routinely performed, these factors are not taken into account in clinical assessments.
\section{Related Work}
The ability to simulate and analyze flow in actual physiological structures, such as patient-specific blood vessels, was made possible by complementary progress in image processing, high-performance computing, and the creation of advanced clinical visualization methods~\cite{bluestein2017utilizing}.
Typically, the structure of interest is segmented from a volume image, relevant features are extracted, and then the CFD simulation can be carried out:
A large body of visualization methods has been developed that enable analyzing the resulting flow fields~\cite{oeltze2019generation}.

\noindent
\textbf{Model Extraction and Pre-processing.} Various techniques for segmentation and surface reconstruction of vascular structures from volume images have been developed.
Traditional techniques such as intensity-based approaches, deformable models~\cite{manniesing2006vessel}, and graph-based methods~\cite{bauer2010segmentation, esneault2010liver} have shown useful results, but they often suffer from inaccuracies due to image artifacts and can be computationally slow~\cite{lesage2009review}.
Recently, machine learning-based methods, particularly convolutional neural networks, have gained popularity for automatic segmentations of image data with noise, artifacts, or different types of atherosclerotic plaque~\cite{liskowski2016segmenting, moeskops2016deep, wu2016deep}.
Different networks for carotid lumen and plaque segmentation have been proposed~\cite{bozkurt2018inverse, cuisenaire2008fully, zhou2021fully}.
%
%
The extracted surfaces are usually further processed to provide additional information, such as landmarks or morphological features.
Particularly relevant are centerline extraction methods that produce a topological model of the vessel tree structure~\cite{saha2016survey}.
%
%
%
A common approach to build a centerline, that is also used in this work, is to connect the origins of maximally inscribed spheres inside the model, which computes the minimal radii of the vessel as a byproduct~\cite{antiga2003computational}.

\noindent
\textbf{Hemodynamic Simulation.} Given a vessel wall model, an internal flow field can be simulated.
The process typically involves many additional steps, including capping the ends of the model, building a volumetric mesh from the surface, and creating profiles for incoming and outgoing flow.
Depending on the complexity level of the required CFD simulation, a numerical solving procedure must be applied that approximates the behavior of continuous functions that model the velocities and pressures occurring in the flow domain.
Lopes et al.~\cite{LOPES2020110019} review hemodynamic simulation approaches applied to carotid artery models.
Sousa et al.~\cite{sousa2016computational} and Guerciotti et al.~\cite{guerciotti2017computational} compared the simulated velocities with measurements from Doppler ultrasonography.
Both show good agreements between the measurement and simulation.
%
%
CFD-derived flow data can further be processed to yield biomechanical parameters that have been shown to play crucial roles in disease progression~\cite{hartman2021definition}.
%

\noindent
\textbf{Blood Flow Visualization.} Analyzing biomedical flow fields is a complex task which can benefit greatly from tailored visualization methods~\cite{vilanova2014visual}.
Oeltze-Jafra et al.~\cite{oeltze2019generation} identify which types of exploration different approaches facilitate, such as overview representation, probing, and contextualization.
Often, multiple coordinated views of the underlying data are combined to facilitate different purposes.
%
%
%
Techniques like flow lenses can be used to probe a user-selected focus region, while spatial context is provided by the vessel wall model~\cite{gasteiger2011flowlens, tominski2016interactive}.
Most approaches focus on exploring flow in a single vascular model.
Eulzer et al.~\cite{eulzer2021visualizing} applied flow visualization methods for analyzing and contextualizing blood flow in the carotids of patients.
Other applications focus largely on flow parameters in aneurysms~\cite{oeltze2014blood, meuschke2016combined, meuschke2017glyph, meuschke2019visual} and the aorta~\cite{angelelli2011straightening, born2013illustrative, Englund2016, Meuschke_2016_EuroVis}.
The comparison of two data sets with internal flow was performed by Meuschke et al.~\cite{Meuschke2022}.
%
%
To the best of our knowledge, there are no frameworks geared for comparative visual exploration of larger sets of hemodynamic simulations, such as patient cohorts.
However, techniques have been proposed that could be used to compare multiple hemodynamic simulation instances.
These are primarily techniques that simplify complex 3D vascular structures to 2D abstractions~\cite{kreiser2018survey}.
Often used abstractions are \textit{vessel maps}~\cite{eulzer2022vessel}, which are fundamentally map-like visualizations of vascular structures.
Vessel maps have been created of the carotid to view surface fields of flow parameters without occlusion~\cite{eulzer2021automatic} and to track plaque development over multiple sonography scanning sessions~\cite{choi2017conformal, choi2020area}.

\section{Task Analysis} \label{sec:task_analysis}
\noindent For the development and testing of our framework, we collaborated with three independent specialists: a neurologist leading a stroke unit (P1, 17 years of practice), a radiologist (P2, 32 years of practice), and a hemodynamic simulation expert (15 years of experience).
At the beginning of the project we conducted multiple focus group interviews with the goal to extract the clinical tasks and requirements a similarity-based blood flow exploration framework would need to address.
The physicians emphasized the potential impact of such a system:
\begin{quote}
    ``\textit{If I could get flow information immediately from the CT, that would be transformative. The blood flow is really ultimately what I'm interested in. If I can get a reasonable prediction of the flow velocity inside a stenosis, for example, that would be highly valuable. A good estimate would already be better than nothing, we do not work with highly exact values anyway.}'' (P1) 
\end{quote}
%
%
They highlighted the significance of a good automatic matching of the models and efficient exploration of similar cases.
``\textit{No physician would go through the database manually. Ideally, the best matches are shown automatically and I can get a peak of what they look like before I read out any specific values} (P2).''
When asked about which flow parameters should be derived from the simulations, the physicians argued that to complement current clinical decision-making the maximal flow velocity should be the primary focus.
It should be possible to probe the maximal velocity in different branches, e.g., inside a stenosis, but also in the CCA, ECA, and after a stenosis.
%
%
Further, P2 noted that studying WSS distributions is increasingly important for understanding stenosis progression.
Regarding the temporal flow dimension, the physicians argued they are essentially only interested in the peak systolic flow, i.e., the point in the cardiac cycle with maximal arterial flow.
In some cases, the systolic flow velocity is also compared against the diastolic (lowest) flow.
Lastly, we stress that the applicability of the proposed framework depends highly on the efficiency of the model extraction.
To compute any similarity metric between a new case and a pre-computed CFD simulation, a geometric vessel model must be available.
We quickly determined that using CTA imaging is the most sensible starting point, as CTAs are acquired routinely in stroke diagnostics, in both preventive and acute cases.
CTA is a comparably fast procedure that images the arterial lumen with a good contrast and is simultaneously used to rule out internal bleeding.
Therefore, to make the system usable in practice, we must integrate an efficient way to generate a geometric model from a CTA volume that we can then compare against the flow database.
In summary, we distilled the following tasks:
\begin{itemize}[parsep=2pt,leftmargin=0.8cm]
    \item[\textbf{T1}] Extract a carotid bifurcation vessel wall model from CTA.
    \item[\textbf{T2}] Find matching models with pre-computed flow fields.
    \item[\textbf{T3}] Visually compare the new model against the selected models.
    \item[\textbf{T4}] Explore relevant flow parameters in the selected models.
    \item[\textbf{T5}] Probe any subregion of the flow fields and extract the maximal local flow velocity.
\end{itemize}

\begin{figure*}[t]
\centering
\includegraphics[width=0.85\textwidth]{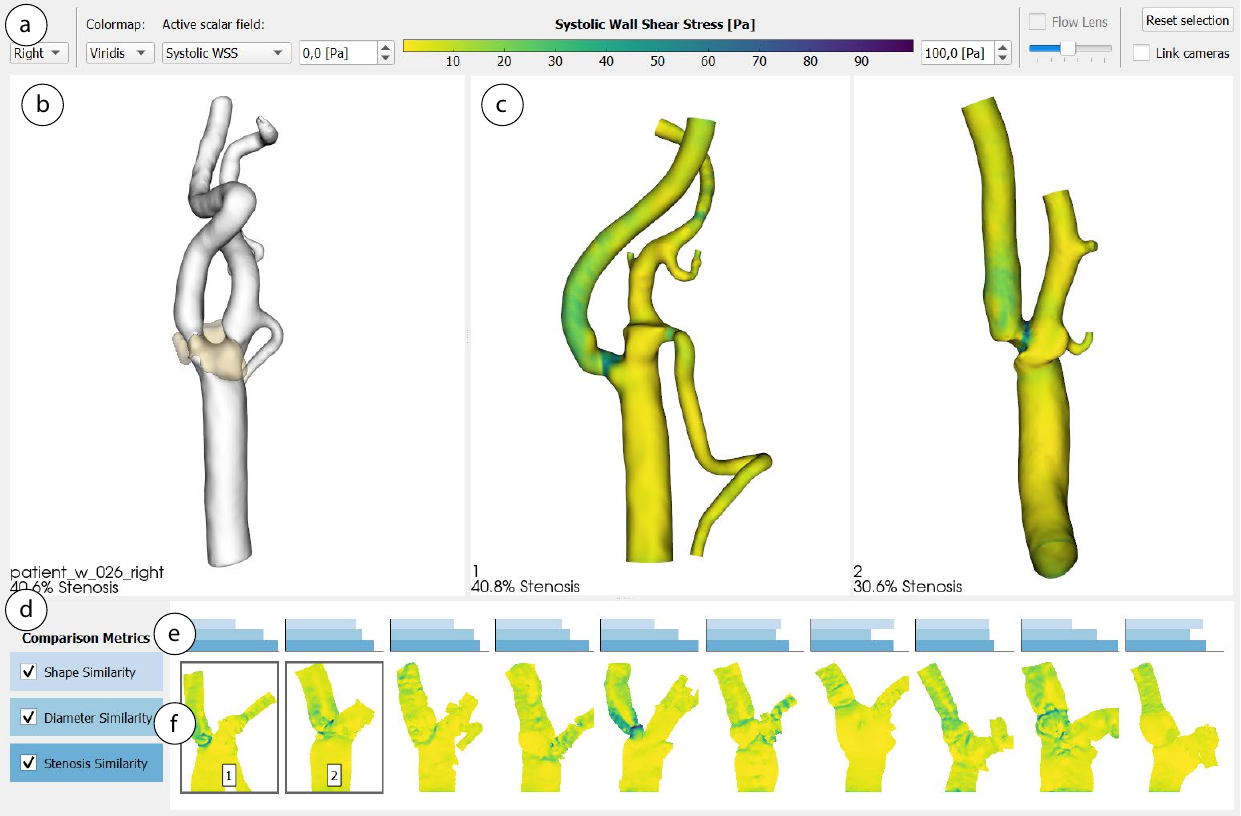}
\caption{The user interface of the visualization framework. (a) Toolbar with parameter selection and color map. (b) 3D view of the vessel surface geometry. (c) 3D views of the selected cases from the flow database. (d) Panel to control which similarity metrics are queried. (e) Bar charts showing how similar a match is regarding each metric. (f) Vessel surface maps giving an overview of the flow parameters.}
\label{fig:overview}
\end{figure*}
\section{Methods}
This work is a first attempt at comparison-based blood flow analysis.
Our primary objective is to test the viability of this approach.
For this, we require an initial database of hemodynamic simulations, which allows us to find geometrically proximate matches to a new carotid bifurcation wall model.
When choosing the size of the database, we must consider the number of instances required to test the viability of the system, but also the feasibility of performing a numerical simulation for each instance.
Studies that employ hemodynamic simulations in patient-specific vascular anatomy typically only focus on very few subjects.
The processes of image data acquisition, model extraction, pre-processing, and numerical simulation incur substantial time and computational costs.
In their systematic review, Lopes et al.~\cite{LOPES2020110019} identified 49 studies on the simulation of blood flow in patient-specific geometries of the carotid artery bifurcation.
Out of these studies, the majority are based on $1$--$5$ bifurcation geometries.
Recently, some larger studies with up to $50$ carotid bifurcations were conducted~\cite{lee2019effect, azar2019geometric}, but these remain exceptions.
While statistical evidence is lacking regarding the number of cases sufficient to cover the anatomical variations of carotid stenosis, we assume that we require more cases than even these larger studies.
Consulting our collaborating physicians, we determined that it would be sensible to include at least 100 variations in the database to be able to reasonably test the approach.
We also need to consider additional overhead for testing, i.e., we need models with simulated flow that are not part of the test database.
Consequently, we have factored in a 20\% overhead and extracted 120 individual carotid bifurcation models from CTA volumes. 
%
%
From the radiological report of each case, we have prior knowledge regarding the severity of the ICA constriction.
We chose cases with ICA stenosis close to the BP and aimed to cover the full spectrum of this type of stenosis.
Therefore, we systematically selected an approximately equal number of cases with no/mild (0--49\%), moderate (50--69\%), and severe (70--99\%) stenosis.
Generally, we extracted both the left and right carotid in each scan, however, we excluded arteries with full occlusion (100\% stenosis), as these vessel are not discernible in the CTA.
Each extraction covers 80\,mm in height with the BP approximately at the center.
This section contains the carotid bifurcation and any nearby stenoses.
We extract the vessel lumen, therefore, the surfaces model the inner vessel wall.
The wall thickness is not reliably detectable in CTA.

\subsection{Creating a Database of Carotid Flow Models} \label{sec:flow_sim}
%
We performed flow simulations on the segmented vessels, including the full extracted bifurcation domain.
Only the tips of the vessel branches were trimmed to define cross-sectional areas for the flow inlets and outlets.
Each model has one flow inlet at the CCA and one outlet at the ICA.
The ECA splits into an arbitrary number of sub-branches, for each of which we modelled an outlet.
The blood flow in the carotid artery tree was simulated by solving the underlying incompressible Navier-Stokes equations for Newtonian fluids.
This was achieved in the FEM (Finite-Element-Method) software COMSOL $6.0$~\cite{comsolmanual}.
The fluid-flow problem was solved on a rigid domain with constant fluid density $\rho_f=1000$\,kg/m$^3$~\cite{vitello2015blood}.
We set the  viscosity  to $\mu= 0.00345$\,Pa\,$\cdot$\,s, which is typically used for blood as a Newtonian fluid as it is the lower viscosity limit of the shear-thinning models $\mu_\infty$~\cite{mendieta2020importance}.
At the inflow, we prescribe a cyclic velocity function that mimics a cardiac cycle.
We use an instantaneous inflow velocity boundary condition corresponding to a typical CCA temporal flow rate profile, see Richter et al.~\cite{richter2023}.
The mean flow is 432\,ml/min.
This results in an average inflow velocity of $0.22$ m/s, which corresponds to the normal flow velocity measured in the CCA of adults~\cite{lee2013general}.
The no-slip boundary conditions are prescribed on the vessel walls.
The outflows are modeled with a boundary condition that simulates the resistance of the down-stream vessel tree.
At each outflow boundary, we implement a varying pressure condition with a resistance approach to split volume flow through the ICA and ECA with a 70\%:30\% ratio.
As long as a stenosis does not fully block the blood flow, this split is a good approximation~\cite{healthyflowvolume, unhealthyflowvolume}.
This is because a developing stenosis is first compensated through dynamic cerebral autoregulation, which is an important physiological mechanism that maintains constant cerebral blood flow~\cite{panerai2007Cerebral}.
The result is an increase in flow velocity inside the stenosed region.
As Vignon-Clementel et al.~\cite{vignonclementel}, we specify the total vascular resistance $R_{tot}=P_{mean}/Q_{mean}$ with the mean pressure obtained as a combination of systolic and diastolic pressures, $p_{mean}=(2\cdot 80\,\mathrm{mmHg} + 120\,\mathrm{mmHg})/3$ and a mean flow rate of $Q_{mean}\approx 7.2\,\mathrm{ml/s}$.
According to the law of adding resistances in parallel, we get resistances $R_{int}=R_{tot}/0.7 $ and $R_{ext}=R_{tot}/0.3$ for the ICA and ECA, respectively.
Further branching of the ECA distributes the flow volumes according to their cross-sectional outflow areas $A_{ext_i}$ for $i=1,\dots,k$ in relation to the total external outflow area $A_{ext}=\sum_i A_{ext_i}$. 
After an initial cardiac cycle, the second cycle is used to evaluate the flow field parameters to mitigate initialization errors.
We derive velocity, pressure, and the risk parameter WSS.
Our volume meshes vary from $9 \cdot 10^5$ to $45 \cdot 10^5$ elements with an average around $20 \cdot 10^5$.
The meshes are mostly composed of tetrahedral elements and include two boundary layers of prism elements to adequately resolve the velocity gradients near the wall.
We apply linear finite elements representing the piece-wise linear numerical solution for both the velocity and pressure field.
For equal polynomial order for velocity and pressure representations, the FEM-method is stabilized by streamline and crosswind diffusion.
We use an adaptive time-stepping scheme with a maximum time step constraint of $0.01$\,s.
The mean simulation time for one artery, excluding pre- and post-processing, is 4 hours 50 minutes on an Intel\textregistered \ Xeon\textregistered \ Gold 5122 CPU at 3.60 GHz with 4 cores.
The total pure computation time for all arteries was 34 days and 9 hours.
While the time requirements for simulation could be reduced by faster hardware, the pre- and post-processing are also time-intensive.
Each data set required an expert additionally 1--3 hours to prepare, setup the simulation, and export the results.

\subsection{Extracting Carotid Bifurcation Geometries}
\noindent While the CFD simulation can be externally computed to build the flow database, any new bifurcation geometry must be extracted from CTA imaging within the clinical workflow (\textbf{T1}).
We address \textbf{T1} by building our flow visualization methods as an extension module to the CarotidAnalyzer pipeline~\cite{eulzer2023fully}.
The CarotidAnalyzer provides a user interface for importing CTA volumes in standardized DICOM format.
A specially trained convolutional neural network is used to label the carotid bifurcation lumen and plaque~\cite{Cardoso_MONAI_An_open-source_2022}.
The resulting labels can also be manually adjusted by the user.
Then, surface models are automatically extracted and the pipeline can further be used to compute vessel tree centerlines using the method of Antiga et al.~\cite{antiga2003computational}.
These centerlines also provide diameter information at every point inside the model.
Using this information, the CarotidAnalyzer computes a stenosis degree (0-99\% closure), following the standard guidelines from the North American Symptomatic Carotid Endarterectomy Trial (NASCET)~\cite{NASCET}.
In summary, we use the CarotidAnalyzer to efficiently extract carotid bifurcation surface geometries, centerlines, the inner vessel diameter, and derive the stenosis degree.
During testing, we found that after a short introduction users can perform the segmentation and pre-processing steps for a new dataset within 2--10 minutes.
The extraction is executed in a single application, it is fully guided by a user interface, and it requires no knowledge of the underlying algorithms.
The extracted models are automatically imported into our extension module, where they can be instantaneously compared against the flow simulation database.

\subsection{Visualizing the Alignment Space Neighborhood}
\label{sec:visualizing_the_neighborhood}
To visualize different carotid data comparably and measure their similarity, we first align them, as the individual models are arbitrarily positioned.
We register the models such that the bifurcation point $BP=(0,0,0)$ and apply a rotation to align the models along their three major branches, i.e., the CCA, ICA, and ECA.
The rotation applies the optimization approach by Ugray et al.~\cite{Ugray2007}, to minimize the average Hausdorff distance between the models.
The detailed matching algorithm can be found in the supplement.

To facilitate \textbf{T2} we need to query the flow database for matching cases.
First, we derive the geometric similarity $S_g$ between the models.
We define $S_g$ as the remaining Hausdorff distance after registration, i.e., it is a byproduct of the branch alignment.
We found that for finding matching cases, using this similarity of the vessel tree shape alone does not yield ideal results, as it does not incorporate information on the vessel thickness.
Therefore, we use the layout-independent information we gain from the model extraction pipeline as additional similarity measures, namely, the point-wise difference of the inner vessel diameter $S_d$ and the difference of the stenosis degree $S_s$.
As a standard clinical measure, the stenosis degree is an established comparison metric already.
However, $S_s$ alone can be misleading, as different stenosis shapes and locations can produce the same degree.
Therefore, we further use the variation in vessel diameter of the CCA and ICA branches.
We align the diameter profile of any two subjects at the BP and compute their point-wise distance.
As the extracted branches differ in arc length, the longer profile is clamped to match the shorter one when computing $S_d$.
When experimenting with test cases manually, we found that good search results ensue from merging the three metrics.
$S_g$ measures similarity regarding the branching geometry and layout.
$S_d$ measures similarity regarding the vessel diameter and \textit{where} a stenosis is located. 
$S_s$ measures similarity regarding the severeness of the constriction.
To make the values comparable, we normalize each metric linearly to $\bar{S}_g, \bar{S}_d, \bar{S}_s \in [0,1]$.
For each metric, the maximum 1 is a theoretical identical match and the minimum 0 is assigned to the database model with lowest fit.
Given a new input geometry, $\bar{S}_g, \bar{S}_d, \bar{S}_s$ are relative measures of how well a model in the database fulfills each metric.
The normalized metrics can be arbitrarily weighted to derive a total distance between two data sets.
As a starting point for our evaluations, we use an evenly weighted linear combination $\bar{S}_{gds} = 1/3(\bar{S}_g+\bar{S}_d+\bar{S}_s)$, which showed good preliminary results.
A first measure of the performance of this matching is provided in \cref{sec:measures}.

The candidates with highest similarity $\bar{S}_{gds}$ are provided for visual comparison.
We found that only showing a single match may not be the best approach, as the similarity search is not a trivial task.
Different candidates may be relevant, depending on what the user is looking for.
Therefore, we explicitly visualize the subset of the flow database closest to the new model regarding $\bar{S}_{gds}$.
We call this subset the alignment space neighborhood.
As we do not know how many samples are required, the user can choose the neighborhood size arbitrarily.
To show how similar a model is regarding each metric, we visualize a bar chart for each candidate, where each metric $\bar{S}_g$, $\bar{S}_d$, $\bar{S}_s$ is shown by one bar.
We use these bar charts, as they are instantly readable even at a small size.
The axis values are omitted, as the numerical value of each bar is meaningless and only the relative length of each bar is relevant.
We sort the candidates automatically and display them below a view of the newly extracted 3D carotid model, see \cref{fig:overview}~(e).
The candidates are shown with decreasing fit from left to right.
We implement an option to toggle individual metrics on or off, which re-sorts the candidate list.
This allows, for example, to run the following queries:
\textit{Is there a model with the same stenosis degree?
Give me a model with a similar diameter profile.}
We quickly noticed that when displaying the alignment space neighborhood, a visual representation of each candidate's geometry is highly beneficial for navigation.
Providing visual cues about the geometry and flow parameters increases recognizability and accelerates finding the ideal match.
An overview of the flow parameters also allows the efficient identification of different types of models.
For example, the user might want to get a best and worst-case comparison.

When visualizing ten or more 3D vascular models, scalability becomes an issue.
Visually assessing multiple complex 3D renderings is problematic, as small depictions become hard to read, 3D interaction would be required to view the full surface, and self-occlusion of the branches obstructs the view.
To address these shortcomings of 3D renderings, vessel maps have been shown to be an effective tool.
They can provide an instant overview of surface fields, require no interaction, and allow the comparison of many instances since they remain readable even at small scales~\cite{eulzer2022vessel}.
Therefore, we show flattened maps of the vessel wall surfaces in the candidates view.
We computed a surface map of every instance in the flow database and visualize these maps below the corresponding bar chart, see \cref{fig:overview}~(f).
We use the method of Eulzer et al.~\cite{eulzer2021automatic} to compute the maps, which is a global surface parameterization technique aimed at creating a single patch for vessel trees.
The method automatically cuts and flattens tree-like surface meshes with an arbitrary number of branches.
It uses vessel cuts that are positioned on one side of the cylindrical vessel branches, enabling intuitive 2D views of the unrolled vessel wall.
The approach preserves vertex and connectivity information, facilitating mapping of data fields defined on mesh vertices.
It minimizes area distortion while maintaining proximity to the original branch layout.
The area optimization is also useful to identify stenosis candidates.
%
%
%
We use the resulting maps to show the flow-derived surface parameters.
We pre-render the maps as $1000 \times 1000$ pixel textures, which means they can be quickly loaded and displayed.
The textures encode a scalar value of each surface field.
The active surface parameter, the colormap, and the visualized value range are chosen by the user.
Any perceptually uniform colorCET colormap~\cite{kovesi2015good} can be applied.
The user can zoom into the candidate list, to display fewer candidates and read details on the maps, and zoom out to show more candidates.
How many candidates are sensible depends on the screen size.
With the maps, specific queries can be made at a glance.
For example, during testing we encountered the following queries:
\textit{Is there a similar model that shows signs of turbulence? Give me a similar model with high WSS at the bifurcation. How fast is the flow in a slightly worse stenosis?}

\subsection{Comparative Visualization of Multiple Flow Models} \label{sec:flow_vis}
To enable \textbf{T3}, we need to facilitate a comparison of the new case with any selected matching models from the flow database.
Visual comparison can be implemented by a superimposition of instances, juxtaposition, interchange, or explicit encoding~\cite{kim2017}.
In our case, the complex vascular geometry excludes superimposition as an option, as the view would become cluttered.
We do not know how many comparisons are required and the user should be able to inspect each geometry individually.
Therefore, we use a juxtaposition to enable a comparison of the 3D geometries and flow fields.
By clicking on a vessel map, a view of the corresponding 3D model is toggled on or off.
An arbitrary number of models can be displayed, which are automatically laid out in a grid beside the inspected case, see \cref{fig:overview}~(c).
%
%
An apparent disadvantage of these side-by-side views is that the 3D interaction becomes cumbersome for multiple models.
For a more fluent interaction, we integrated a \textit{synchronized} juxtaposition that can be enabled on demand.
In the synchronized view, we transform the displayed models into the alignment space where they are registered.
Then, the camera is synchronized between all views.
The flow information is mapped on the models to address \textbf{T4}.
The same colormap that is set for the vessel maps is shared across all depicted surface models.
%
%
If instead of a surface field a velocity field is chosen, we render streamlines of the flow and contextualize them with the backside of the vessel wall.
%
%
We chose this streamline encoding, because primarily the single time step of peak systolic velocity is relevant to the users and the streamlines reveal features of interest, such as poststenotic vortices.
For the integration, we use the 5th order Runge-Kutta method~\cite{kutta1901beitrag, runge1895numerische}.
The streamline seeds are placed in 300 random cells of each carotid volume mesh to distribute them across the mesh.
The streamline geometry is pre-computed and cached for each case, therefore, they can be immediately shown once a case is chosen for display.

\begin{figure}[t]
\centering
\includegraphics[width=1.0\linewidth]{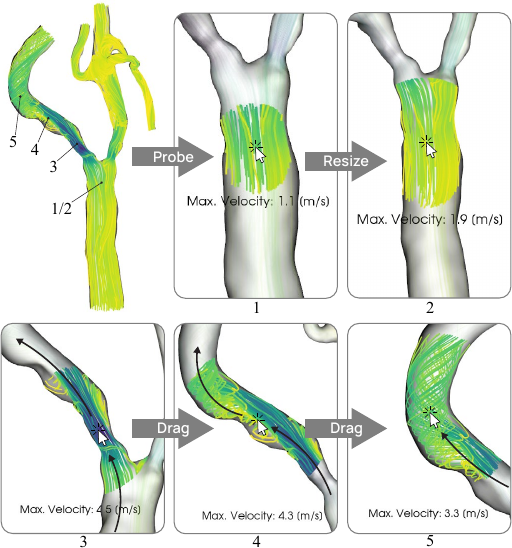}
\caption{A probe lens can be interactively dragged over the vessel to focus on the flow in a specific region. Flow properties, such as the maximal velocity within the selection, are extracted automatically.}
\label{fig:probing}
\end{figure}

Lastly, \textbf{T5} requires functionality to probe the flow field at user-defined locations and extract the maximal local flow velocity.
We implemented a type of flow lens, which can be activated when the flow velocity is displayed.
When the flow lens is enabled, we shoot a ray into the scene of the active viewport, originating from the cursor position on the screen.
We determine the first hit position of the viewport's vascular model or, if no hit is found, the closest surface point to the ray.
At this point, we create a sphere with a user-adjustable radius.
The sphere is used as an implicit function to filter the streamlines and we crop the line geometry outside the sphere.
We filter the actual 3D geometry, not the image space, which allows us to accurately determine the flow velocities inside this region.
The extracted value is displayed next to the probed region, see \cref{fig:probing}.
The probe can be continuously moved by dragging the cursor, allowing smooth exploration of the flow field and determining the flow velocity in all required regions, e.g., inside the stenosis, but also in the CCA or behind the stenosis.
\section{Evaluation}
We evaluated the numerical soundness of the comparison-based flow analysis to get a baseline of how well flow parameters can be predicted with the system.
We also tested the interactive framework with five physicians working in stroke care.

\subsection{Numerical Evaluation} \label{sec:measures}
We want to quantify the current error of the prediction when compared to performing a case-specific CFD simulation.
For this purpose, we applied a standard 80/20 split on the database, leading to 96 remaining models in the prediction database and 24 test instances.
The test cases were randomly selected.
For each test instance, we measured the deviation between the simulation ground truth and the system's prediction.
For the prediction, we use the database geometry with the highest similarity metric $\bar{S}_{gds}$, which for simplicity we will call the \textit{best match}.
Furthermore, to assess the robustness of the similarity metrics, we compare $\bar{S}_{gds}$ against choosing the best match based on the smallest Hausdorff distance, which is a widely applied standard for matching geometric objects~\cite{aydin2021usage, ryu2021efficient}.
For each match, we compute how the flow properties differ between the ground truth and the prediction.
For our comparisons, we focus on flow parameters relevant to clinical contexts as well as flow field analysis.
We determined the prediction error regarding velocity, vorticity, WSS, and pressure.
The systolic peak velocity is highly relevant as it is used for stenosis grading.
The occurrence of poststenotic turbulent or vortex-like flow is also used as a secondary marker in clinical stenosis grading~\cite{aimw2020s3}.
We measure the vorticity as a property of the flow structure.
Note, however, that vorticity has limited expressiveness, as it is not only influenced by vortices but also the velocity and shear flow near vessel walls.
WSS and pressure measure the forces acting on the vessel walls and have been linked to plaque formation in the carotids~\cite{li2019hemodynamic}.

Comparing the flow properties of a best match to the ground truth is not a straightforward task, since the wall geometry and layout of the test instance can differ from the match, even after aligning the models.
A match may be very similar in terms of the diameter and stenosis degree, leading to a high similarity score, but have branches of a different length and/or layout.
This means a direct comparison of the flow fields is impractical.
To enable a sensible comparison, we use predetermined samples of the flow field at the clinically most significant and uniformly recognizable landmarks: inside the CCA close to the bifurcation, inside the ICA stenosis, and shortly after the stenosis.
These sample positions are indicated in \cref{fig:probing}, positions 2, 3, 4.
They are the relevant locations examined in carotid ultrasound diagnostics.
The hemodynamic parameters at the sample location are averaged within a 5\;mm radius to prevent noise from affecting the measurement.
We also measure the peak velocity, i.e., the maximum value within each sample region.
For every test instance, we sample the three landmarks at the peak systolic and end-diastolic time points, as these are the two time points used in clinical assessments.
By computing the absolute deviation for each sample location and time point between the case-specific simulation and the prediction, we determine the prediction error.

\subsection{Structured Interviews} \label{sec:interviews}
\noindent To validate if the tasks \textbf{T1}--\textbf{T5} can be performed, we conducted qualitative interviews with five physicians (P1--P5; two female, three male; ages 30--63).
They have 17, 32, 13, 9, and 15 years of working experience in clinics for neurology or radiology.
P1 and P2 are co-authors of this work, they have also been part in earlier discussions and for the task analysis.
Also, they acquired the CTA data and validated the lumen/plaque segmentations we used to create the flow database.
None of the participants had used the developed software framework before.

We conducted individual in-person interviews lasting about 60 minutes, where we first acquainted the participants with the concept of the similarity-based hemodynamics analysis.
Then, we demonstrated how a model can be extracted from CTA using the CarotidAnalyzer pipeline and how the wall geometry can be viewed in our extension module.
We explained and showed all exploration features, e.g., how the neighborhood can be filtered, how candidates can be selected, and how the flow parameters can be analyzed.
After the introduction, the participants were asked to analyze three new CTAs, which were not included in the flow database.
One represented a healthy patient, one contained a mild, and one a severe stenosis.
We asked the participants to extract one geometry from each CTA volume (\textbf{T1}), find one or multiple matching flow models (\textbf{T2}), and compare them to the extracted geometry (\textbf{T3}).
We further asked which flow parameters they could gather from their selection (\textbf{T4}, \textbf{T5}) and if these confirmed their expectations.
During these case studies, we employed a think-aloud protocol.
We followed the case studies with a questionnaire, where we asked questions regarding the fulfillment of each task and regarding the comprehensibility and usefulness of each visual encoding (vessel maps, bar charts, color maps, streamlines, probe).
The participants rated each question on a five-point Likert scale ($--$, $-$, $\circ$, $+$, $++$) and we noted down verbal comments regarding each item.

\begin{figure*}[t]
\centering%
\includegraphics[width=1.0\textwidth]{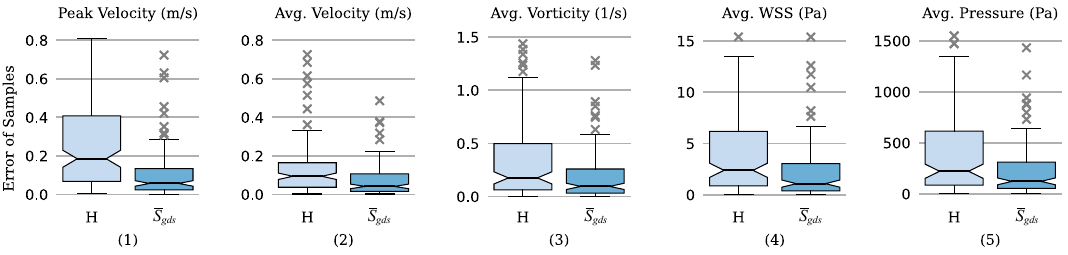}%
\caption{Boxplots of the error between the ground truth simulation and prediction. H: The prediction was selected based on the Hausdorff distance. $\bar{S}_{gds}$: The prediction was selected based on our combined similarity metric.}%
\label{fig:deviations_boxplots}%
\end{figure*}

\section{Results and Discussion}
The results of the numerical evaluation are shown in \cref{fig:deviations_boxplots}.
Exact values are provided in the supplement.
For the majority of the samples, the predictions are highly accurate.
For the peak velocity, we achieve a median error of $0.06$~m/s (using the $\bar{S}_{gds}$ matching metric).
For the average velocity, the median error is $0.04$~m/s.
Our mean errors are $0.12$~m/s (peak velocity) and $0.08$~m/s (average velocity).
The magnitude of these velocity errors can be better interpreted by looking at the clinical measurement variability.
The mean interobserver variabilities for ultrasound assessments of the CCA and ICA peak velocities were observed to range from $-0.08$ to $0.14$ m/s~\cite{mead2000variability}.
%
%
The limits of agreement for individual patients were found to be [$-0.68$, $0.85$]~m/s~\cite{mead2000variability}.
Our limits of agreement for individual samples are [$-0.72$, $0.82$]~m/s (peak velocity) and [$-0.79$, $0.28$]~m/s (average velocity).
This means, for the clinically relevant flow velocity, our prediction errors are within the currently tolerated limits.
The system showed to be able to find similar cases and predict flow velocities with errors comparable to the disagreement between physicians when performing velocity measurements.

Generally, the median error of all assessed parameters is low with regard to the typically observed value range.
The mean error, however, in all cases is higher than the median error, indicating a skewed distribution due to some predictions with disproportionately larger errors.
This assumption can be verified by looking at the boxplots of the error distributions, see \cref{fig:deviations_boxplots}.
For all flow parameters, at least half of the samples only show small errors close to 0.
However, there are outliers with a larger error which skew the distribution.
25\% of the peak velocity samples and 16\% of the average velocity samples show an error larger than the observed interobserver variability of $0.14$ m/s.
These cases seem to be insufficiently covered by the pre-computed flow fields, leading to the conclusion that the database will need to be extended.

Lastly, we observe that the combined similarity metric $\bar{S}_{gds}$ is consistently better at choosing a prediction model than using the Hausdorff metric (H).
The median error is significantly lower for all tested parameters and the spread of the error distribution is smaller.
These results indicate that the relevant flow parameters are more dependent on the CCA/ICA vessel diameter, stenosis location, and stenosis degree than on the shape or layout of the vessel branches.
A core observation of this study is that the clinically relevant blood flow parameters can be predicted by a vessel model with a similar stenosis (regarding location, length, and diameter), even if the branch layout is different and the number of modelled ECA branches varies.
Our proposed algorithm to build the alignment space appears to accurately match cases with similar flow in the relevant areas, i.e., inside the ICA and potential stenoses, even if no model exists that has a globally equivalent surface geometry.
This observation may aid future research regarding comparison-based flow analysis, as it shows that the database does not need to cover all potential geometric variations of the target anatomy but can be constrained to focus on variations of the crucial features, such as the stenosis position and shape.

\subsection{Interview Results}
%
We report on frequent comments and feedback from the interviews.
The overall impressions of the participants were highly positive.
The processing pipeline was well received and the model extraction (\textbf{T1}) was deemed efficient enough to be applicable.
\begin{quote}
``\textit{I highly appreciate that the software is usable as is. It is very well integrated. I can import a DICOM image and very quickly get an impression of the internal flow field.}'' (P3)
\end{quote}
The participants were generally able to find similar enough comparison cases (\textbf{T2}).
P4 said, ``\textit{Considering you have about 100 models, I am surprised how well the pairing works. I think for some we could get even better results, but that is just a question of extending the database.}''
The comparison metrics were understood and all experts noted they found the metrics sensible.
P1 emphasized, ``\textit{Especially the diameter makes total sense. I think the exact progression of the branches is not so important, so it's nice that I can turn that filter [geometric similarity] off.}''
We observed that typically one or two and rarely more than four comparison cases were used simultaneously.
Two participants noted they would only want to see the closest case and that ``\textit{the less interaction is required the better}'' (P2).
However, opinions on this matter diverged.
P5 noted, ``\textit{It highly depends on the variance. If I can get one really good match it is sufficient. But otherwise, I definitely need three or more similar cases for comparison.}''
Also, the option of comparing a slightly worse stenosis to assess the possible progression was mentioned.

Regarding the clinical relevance of the shown parameters, we gathered the impression that as long as no clinical guidelines exist, using WSS is only of supplemental importance.
%
%
All experts agreed that the peak flow velocity is currently of the highest clinical relevance.
P2 argued, ''\textit{I would still definitely keep the other parameters, especially WSS is increasing in importance. Having it available is never a drawback.}''
When asked about the temporal resolution, every interviewed expert underlined that they are almost exclusively interested in the peak systolic flow and would not see the upside of including more timesteps.
Furthermore, all participants stated that the bar charts used to depict the model similarity are highly useful to judge the appropriateness of a case.
The vessel maps were also positively received as a way to get an overview of the displayed candidates: P5 said, ``\textit{I used the maps extensively to choose a model that reflects my presumptions about the flow. I want to see them and make the decision which model should be compared.}''
When assessing the 3D models (\textbf{T3}, \textbf{T4}), the participants mentioned that the streamlines were helpful to judge if turbulence occurred.
All five physicians remarked that the ability to synchronize the model orientations is highly beneficial.
For example, P3 said, ``\textit{The linked perspective is extremely helpful. I first struggled to align all the cases manually, it is so much faster with the linking.}''
They also stated that they found the probing interaction (\textbf{T5}) very useful and intuitive.
P2 said, ``\textit{The probe is very helpful, also that I can change its size. It's like an enhanced ultrasound, I can peek into every branch and get a quick reading of the velocity.}''
After the interview, three participants expressed immediate interest to deploy the tool and proposed to perform a validation study to compare the results against the measured flow.
Notably, P4 described a further application scenario:
\begin{quote}
``\textit{Every day, I have to decide if we transfer stroke patients from smaller clinics to the university hospital. They cannot do a Doppler ultrasound there, I have to judge the stenoses on the CTA alone. This is also a question of resource management, we cannot transfer everyone and very often we transfer unnecessarily, most patients do not need surgery. If I can get a good prediction of the flow from the CT images, that would vastly improve these decisions. I would use this on a daily basis.}'' (P4)
\end{quote}
%

\subsection{Limitations}
In many of the test instances of the numerical evaluation, we observed that the approach produces suitable results.
If vessel wall models with high similarity in the crucial areas are available in the database, the framework accurately predicts the flow properties of a case-specific CFD simulation at the clinically relevant landmarks. 
However, we also observed test instances with a higher prediction error.
These instances are visible mainly as outliers in the distribution of the sample errors.
Unusually large errors occurred if no model with a high similarity metric was available.
Keeping in mind that the database we used during the testing only involved 96 models, we still consider the results of the numerical evaluation a positive outcome.
We demonstrate that with the existing pre-simulated cases, we can already predict the clinically relevant flow parameters in a large portion of carotid stenosis cases and achieve velocity errors similar to the interobserver variability of physicians.
However, for a usable application, the current database needs to be extended.
This extension should not only increase the database size but also improve the data variation.
When selecting CTA scans, we only differentiated them by stenosis degree.
To fully sample the shape space of ICA stenosis, other geometric factors should be considered.
This includes the stenosis position, length, and also local properties such as the regularity of the internal wall structure.
We also noticed that since the velocity does not increase linearly with the stenosis degree, we require further artery models with severe stenoses to achieve more precise results in these edge cases.
One option to improve the database size and variation would be to introduce artificial modulations to the data.
By varying the position and degree of stenoses, their shape space could be systematically sampled.
With artificial geometries, however, it could be difficult to reproduce exactly how stenoses develop.
In general, more testing will be required to form a better understanding of the relation between the database size and quality of results.

While the developed combined metric $\bar{S}_{gds}$ showed better matching results than the Hausdorff distance, we did not test other weightings or other measures.
It would be conceivable to use more refined approaches.
For instance, when comparing the vessel diameters, $\bar{S}_d$ is based on a fixed alignment, which could benefit from techniques like the Fréchet distance that would optimize the alignment of the diameter profiles.
When the database of flow models grows and geometric features become more nuanced, the discriminative power of the metrics may also be insufficient.
Overcoming this issue would require an adaptation of the metrics that takes the actual surface geometry more into account.

CFD necessarily needs to employ simplifications.
Standard boundary conditions model realistic but averaged parameters that might not apply to every patient.
Some CFD studies use patient-specific boundary conditions, such as inflow velocities that were recorded using Doppler sonography.
Using such boundary conditions requires a measure first, which defeats a substantial advantage of computed hemodynamics -- dynamic flow data can already be generated from static image data, without additional procedures.
Such simulated flow provides an estimate of important hemodynamic parameters from a single CTA scan but necessarily requires the use of standardized boundary conditions.
Many of the resulting parameters, e.g., the peak velocity and the occurrence of poststenotic turbulence, are key factors for stenosis evaluation but cannot be directly derived from CTA.
An estimate of the blood flow can already provide a substantial benefit, especially where measurements have not been or cannot be conducted, for example, if a stroke patient was delivered to a hospital where Doppler ultrasonography is not performed.
Ultimately, the accuracy of the simulations determines their viability.
%
%
First studies have shown that velocities measured with Doppler sonography can be reproduced with CFD~\cite{guerciotti2017computational, sousa2016computational}.
However, larger studies with more subjects are pending.
While many studies exist that link WSS to plaque progression~\cite{hartman2021definition}, further long-time studies are necessary to make flow-derived parameters part of clinical guidelines.
The feasibility of extensive clinical CFD studies depends highly on the applicability of hemodynamic simulation in real-world contexts.
The method described in this work addresses the latter, specifically the challenging time and resource-intensive nature of these simulations.
Our approach should be seen as a complement to research targeting the validity of hemodynamic simulation.

\subsection{Opportunities and Future Work}
Two interviewed physicians independently mentioned that the flow parameters of multiple models could be interpolated.
This could provide more accurate values and projecting them on the new geometry would make them directly readable on the extracted vessel.
However, this approach might also be riskier than the purely comparison-based analysis, as the interpolation could be misleading if no good match is available.
Similarly, first studies have shown that hemodynamic surface parameters like WSS can be predicted using convolutional neural networks~\cite{sun2020zernet}.
The disadvantage compared to our approach is that the prediction model is obscure to the observer, i.e., it is not clear where the information comes from.
In our case, the user can immediately see if no reasonably similar vessel geometry is found and if the similarity metrics are low, instead of getting a potentially misleading prediction.

In the future, we intend to extend both, the flow database and the simulation domain.
The latter was proposed by one study participant, who noted that extending the region to include intracranial vessels would be highly beneficial.
Measuring intracranial flow velocities with Doppler ultrasonography is extremely challenging due to the distortion of the ultrasound signal by the skull.
Yet, it is still a widely adopted procedure, as non-invasive alternatives are missing.
Likewise, we plan to apply the similarity-based flow analysis to aneurysms, for many of which only static imaging is available.
Similarity-based analysis could also help to find similar cases with other information than flow.
For instance, from an aneurysm database, geometrically close cases could be queried for which a rupture status is known.
This would allow to derive a geometry-based rupture probability.
\section{Conclusion}
We presented the first approach for visual analysis of hemodynamics using approximated flow in morphologically similar models.
While we focus our implementation and testing on blood flow in carotid bifurcations, the concept could equally be transferred to other transport problems, like intracranial circulation or the hemodynamics of aneurysms.
We propose a combination of similarity metrics, which represent the distances between a new vessel wall model and models in a database.
For the models in the database, we pre-computed flow fields, relevant flow-derived parameters, and velocity streamlines.
Given a new carotid bifurcation, the most similar geometries with computed flow can be queried from the database and visualized, providing an instantaneous approximation of the blood flow.
We display an overview of fitting candidates using map-like visualization, and facilitate simultaneous exploration and comparison of flow in selected models.
We integrate our framework as an extension to the CarotidAnalyzer~\cite{eulzer2023fully}, which provides an efficient toolset for the extraction of carotid bifurcation models from CTA images.
We show that for models in the database that are geometrically close, clinically relevant flow parameters, like the peak systolic velocity, are highly similar.
The user study confirmed that the similarity-based flow analysis could benefit critical clinical processes, which demand fast access to relevant parameters such as flow velocity.
In conclusion, we investigated a possibility to combat one of the strongest weaknesses of CFD in medicine -- due to the time-consuming and resource-intensive nature of hemodynamic simulations, they are difficult to directly apply in clinical workflows.
Our hope is that the insights gathered in this work will support efforts to make hemodynamic simulations more accessible and useful for clinical decision-making.

\bibliographystyle{eg-alpha-doi} 
\bibliography{FlowComp.bib}       

\newcommand{\etalchar}[1]{$^{#1}$}
\begin{thebibliography}{\uppercase{MWvdV{\etalchar{*}}16}}

\bibitem[AEIR03]{antiga2003computational}
\textsc{Antiga L., Ene-Iordache B., Remuzzi A.}:
\newblock Computational geometry for patient-specific reconstruction and
  meshing of blood vessels from {MR} and {CT} angiography.
\newblock \emph{{IEEE} Transactions on Medical Imaging 22}, 5 (may 2003),
  674--684.
\newblock \href {https://doi.org/10.1109/tmi.2003.812261}
  {\path{doi:10.1109/tmi.2003.812261}}.

\bibitem[AGG{\etalchar{*}}20]{aimw2020s3}
\textsc{{Association of the Scientific Medical Societies of Germany}, {German
  Vascular Society}, {German Neurology Society}, et~al.}:
\newblock S3 guideline on diagnosis, treatment, and aftercare of extracranial
  carotid stenosis, 2020.
\newblock URL: \url{https://register.awmf.org/de/leitlinien/detail/004-028}.

\bibitem[AH11]{angelelli2011straightening}
\textsc{Angelelli P., Hauser H.}:
\newblock Straightening tubular flow for side-by-side visualization.
\newblock \emph{IEEE Transactions on Visualization and Computer Graphics 17},
  12 (2011), 2063--2070.
\newblock \href {https://doi.org/10.1109/TVCG.2011.235}
  {\path{doi:10.1109/TVCG.2011.235}}.

\bibitem[ATD{\etalchar{*}}19]{azar2019geometric}
\textsc{Azar D., Torres W.~M., Davis L.~A., Shaw T., Eberth J.~F., Kolachalama
  V.~B., Lessner S.~M., Shazly T.}:
\newblock Geometric determinants of local hemodynamics in severe carotid artery
  stenosis.
\newblock \emph{Computers in Biology and Medicine 114} (2019), 103436.
\newblock \href {https://doi.org/10.1016/j.compbiomed.2019.103436}
  {\path{doi:10.1016/j.compbiomed.2019.103436}}.

\bibitem[ATH{\etalchar{*}}21]{aydin2021usage}
\textsc{Aydin O.~U., Taha A.~A., Hilbert A., Khalil A.~A., Galinovic I.,
  Fiebach J.~B., Frey D., Madai V.~I.}:
\newblock On the usage of average hausdorff distance for segmentation
  performance assessment: hidden error when used for ranking.
\newblock \emph{European Radiology Experimental 5}, 1 (jan 2021).
\newblock \href {https://doi.org/10.1186/s41747-020-00200-2}
  {\path{doi:10.1186/s41747-020-00200-2}}.

\bibitem[BKS18]{bozkurt2018inverse}
\textsc{Bozkurt F., Köse C., Sar{\i} A.}:
\newblock An inverse approach for automatic segmentation of carotid and
  vertebral arteries in {CTA}.
\newblock \emph{Expert Systems with Applications 93} (mar 2018), 358--375.
\newblock \href {https://doi.org/10.1016/j.eswa.2017.10.041}
  {\path{doi:10.1016/j.eswa.2017.10.041}}.

\bibitem[Blu17]{bluestein2017utilizing}
\textsc{Bluestein D.}:
\newblock Utilizing computational fluid dynamics in cardiovascular engineering
  and medicine - what you need to know. {Its} translation to the
  clinic/bedside.
\newblock \emph{Artificial Organs 41}, 2 (feb 2017), 117--121.
\newblock \href {https://doi.org/10.1111/aor.12914}
  {\path{doi:10.1111/aor.12914}}.

\bibitem[BMGS13]{born2013illustrative}
\textsc{Born S., Markl M., Gutberlet M., Scheuermann G.}:
\newblock Illustrative visualization of cardiac and aortic blood flow from {4D}
  {MRI} data.
\newblock In \emph{IEEE Pacific Visualization Symposium} (2013), pp.~129--136.

\bibitem[BPS{\etalchar{*}}10]{bauer2010segmentation}
\textsc{Bauer C., Pock T., Sorantin E., Bischof H., Beichel R.}:
\newblock Segmentation of interwoven {3D} tubular tree structures utilizing
  shape priors and graph cuts.
\newblock \emph{Medical Image Analysis 14}, 2 (apr 2010), 172--184.
\newblock \href {https://doi.org/10.1016/j.media.2009.11.003}
  {\path{doi:10.1016/j.media.2009.11.003}}.

\bibitem[CCLC17]{choi2017conformal}
\textsc{Choi G. P.~T., Chen Y., Lui L.~M., Chiu B.}:
\newblock Conformal mapping of carotid vessel wall and plaque thickness
  measured from {3D} ultrasound images.
\newblock \emph{Medical {\&} Biological Engineering {\&} Computing 55}, 12 (jun
  2017), 2183--2195.
\newblock \href {https://doi.org/10.1007/s11517-017-1656-4}
  {\path{doi:10.1007/s11517-017-1656-4}}.

\bibitem[CCR20]{choi2020area}
\textsc{Choi G. P.~T., Chiu B., Rycroft C.~H.}:
\newblock Area-preserving mapping of {3D} carotid ultrasound images using
  density-equalizing reference map.
\newblock \emph{{IEEE} Transactions on Biomedical Engineering 67}, 9 (sep
  2020), 2507--2517.
\newblock \href {https://doi.org/10.1109/tbme.2019.2963783}
  {\path{doi:10.1109/tbme.2019.2963783}}.

\bibitem[CL13]{caballero2013review}
\textsc{Caballero A., La{\'\i}n S.}:
\newblock A review on computational fluid dynamics modelling in human thoracic
  aorta.
\newblock \emph{Cardiovascular Engineering and Technology 4}, 2 (2013),
  103--130.
\newblock \href {https://doi.org/10.1002/mrm.1910400207}
  {\path{doi:10.1002/mrm.1910400207}}.

\bibitem[CLB22]{Cardoso_MONAI_An_open-source_2022}
\textsc{Cardoso M.~J., Li W., Brown R.}:
\newblock {MONAI: An open-source framework for deep learning in healthcare}.
\newblock \href {https://doi.org/https://doi.org/10.48550/arXiv.2211.02701}
  {\path{doi:https://doi.org/10.48550/arXiv.2211.02701}}.

\bibitem[{COM}23]{comsolmanual}
\textsc{{COMSOL Multiphysics}}:
\newblock Online at: \url{https://doc.comsol.com/6.0/docserver/}, {accessed
  March 2023}.

\bibitem[CVOA08]{cuisenaire2008fully}
\textsc{Cuisenaire O., Virmani S., Olszewski M.~E., Ardon R.}:
\newblock Fully automated segmentation of carotid and vertebral arteries from
  contrast enhanced {CTA}.
\newblock In \emph{{SPIE} Proceedings} (mar 2008), Reinhardt J.~M., Pluim J.
  P.~W., (Eds.), {SPIE}.
\newblock \href {https://doi.org/10.1117/12.770481}
  {\path{doi:10.1117/12.770481}}.

\bibitem[EL23]{CarotidAnalyzer}
\textsc{Eulzer P., Lawonn K.}:
\newblock {Carotid Analyzer}, 2023.
\newblock URL: \url{https://github.com/PepeEulzer/CarotidAnalyzer}.

\bibitem[ELD10]{esneault2010liver}
\textsc{Esneault S., Lafon C., Dillenseger J.-L.}:
\newblock Liver vessels segmentation using a hybrid geometrical moments/graph
  cuts method.
\newblock \emph{{IEEE} Transactions on Biomedical Engineering 57}, 2 (feb
  2010), 276--283.
\newblock \href {https://doi.org/10.1109/tbme.2009.2032161}
  {\path{doi:10.1109/tbme.2009.2032161}}.

\bibitem[EMKL21]{eulzer2021visualizing}
\textsc{Eulzer P., Meuschke M., Klingner C.~M., Lawonn K.}:
\newblock Visualizing carotid blood flow simulations for stroke prevention.
\newblock \emph{Computer Graphics Forum 40}, 3 (jun 2021), 435--446.
\newblock \href {https://doi.org/10.1111/cgf.14319}
  {\path{doi:10.1111/cgf.14319}}.

\bibitem[EMML22]{eulzer2022vessel}
\textsc{Eulzer P., Meuschke M., Mistelbauer G., Lawonn K.}:
\newblock Vessel maps: A survey of map-like visualizations of the
  cardiovascular system.
\newblock \emph{Computer Graphics Forum 41}, 3 (jun 2022), 645--673.
\newblock \href {https://doi.org/10.1111/cgf.14576}
  {\path{doi:10.1111/cgf.14576}}.

\bibitem[ERH16]{Englund2016}
\textsc{Englund R., Ropinski T., Hotz I.}:
\newblock Coherence maps for blood flow exploration.
\newblock In \emph{Eurographics Workshop on Visual Computing for Biology and
  Medicine} (2016), The Eurographics Association.
\newblock \href {https://doi.org/10.2312/vcbm.20161274}
  {\path{doi:10.2312/vcbm.20161274}}.

\bibitem[ERM{\etalchar{*}}21]{eulzer2021automatic}
\textsc{Eulzer P., Richter K., Meuschke M., Hundertmark A., Lawonn K.}:
\newblock Automatic cutting and flattening of carotid artery geometries.
\newblock In \emph{Eurographics Workshop on Visual Computing for Biology and
  Medicine} (2021), The Eurographics Association.
\newblock \href {https://doi.org/10.2312/VCBM.20211347}
  {\path{doi:10.2312/VCBM.20211347}}.

\bibitem[ERP{\etalchar{*}}24]{eulzer2024dataset}
\textsc{Eulzer P., Richter K., Probst T., Hundertmark A., Lawonn K.}:
\newblock A dataset of reconstructed carotid bifurcation lumen and plaque
  models with centerline tree and simulated hemodynamics, 2024.
\newblock \href {https://doi.org/10.5281/ZENODO.7634643}
  {\path{doi:10.5281/ZENODO.7634643}}.

\bibitem[EvDW{\etalchar{*}}23]{eulzer2023fully}
\textsc{Eulzer P., von Deylen F., Wickenh{\"o}fer R., Klingner C.~M., Lawonn
  K.}:
\newblock A fully integrated pipeline for visual carotid morphology analysis.
\newblock \emph{Computer Graphics Forum 42}, 3 (2023), 25--37.
\newblock \href {https://doi.org/10.1111/cgf.14808}
  {\path{doi:10.1111/cgf.14808}}.

\bibitem[FEB{\etalchar{*}}99]{NASCET}
\textsc{Ferguson G.~G., Eliasziw M., Barr H. W.~K., Clagett G.~P., Barnes
  R.~W., Wallace M.~C., Taylor D.~W., Haynes R.~B., Finan J.~W., Hachinski
  V.~C., Barnett H. J.~M.}:
\newblock The north american symptomatic carotid endarterectomy trial.
\newblock \emph{Stroke 30}, 9 (sep 1999), 1751--1758.
\newblock \href {https://doi.org/10.1161/01.str.30.9.1751}
  {\path{doi:10.1161/01.str.30.9.1751}}.

\bibitem[{GBD}19]{burden_of_stroke}
\textsc{{GBD 2016 Stroke Collaborators}}:
\newblock Global, regional, and national burden of stroke, 1990–2016: a
  systematic analysis for the global burden of disease study 2016.
\newblock \emph{The Lancet Neurology 18}, 5 (2019), 439 -- 458.
\newblock \href {https://doi.org/10.1016/S1474-4422(19)30034-1}
  {\path{doi:10.1016/S1474-4422(19)30034-1}}.

\bibitem[GNBP11]{gasteiger2011flowlens}
\textsc{Gasteiger R., Neugebauer M., Beuing O., Preim B.}:
\newblock The flowlens: A focus-and-context visualization approach for
  exploration of blood flow in cerebral aneurysms.
\newblock \emph{IEEE Transactions on Visualization and Computer Graphics 17},
  12 (2011), 2183--2192.
\newblock \href {https://doi.org/10.1109/TVCG.2011.243}
  {\path{doi:10.1109/TVCG.2011.243}}.

\bibitem[Gor19]{gorelick2019global}
\textsc{Gorelick P.~B.}:
\newblock The global burden of stroke: persistent and disabling.
\newblock \emph{The Lancet Neurology 18}, 5 (2019), 417--418.
\newblock \href {https://doi.org/10.1016/S1474-4422(19)30030-4}
  {\path{doi:10.1016/S1474-4422(19)30030-4}}.

\bibitem[GV17]{guerciotti2017computational}
\textsc{Guerciotti B., Vergara C.}:
\newblock Computational comparison between newtonian and non-newtonian blood
  rheologies in stenotic vessels.
\newblock In \emph{Biomedical Technology}. Springer International Publishing,
  aug 2017, pp.~169--183.
\newblock \href {https://doi.org/10.1007/978-3-319-59548-1_10}
  {\path{doi:10.1007/978-3-319-59548-1_10}}.

\bibitem[HNG{\etalchar{*}}21]{hartman2021definition}
\textsc{Hartman E. M.~J., Nisco G.~D., Gijsen F. J.~H., Korteland S.-A.,
  van~der Steen A. F.~W., Daemen J., Wentzel J.~J.}:
\newblock The definition of low wall shear stress and its effect on plaque
  progression estimation in human coronary arteries.
\newblock \emph{Scientific Reports 11}, 1 (nov 2021).
\newblock \href {https://doi.org/10.1038/s41598-021-01232-3}
  {\path{doi:10.1038/s41598-021-01232-3}}.

\bibitem[HTW{\etalchar{*}}09]{halm2009risk}
\textsc{Halm E.~A., Tuhrim S., Wang J.~J., Rockman C., Riles T.~S., Chassin
  M.~R.}:
\newblock Risk factors for perioperative death and stroke after carotid
  endarterectomy: results of the {New York} carotid artery surgery study.
\newblock \emph{Stroke 40}, 1 (2009), 221--229.
\newblock \href {https://doi.org/10.1161/STROKEAHA.108.524785}
  {\path{doi:10.1161/STROKEAHA.108.524785}}.

\bibitem[KCK17]{kim2017}
\textsc{Kim K., Carlis J.~V., Keefe D.~F.}:
\newblock Comparison techniques utilized in spatial {3D} and {4D} data
  visualizations: A survey and future directions.
\newblock \emph{Computers \& Graphics 67} (2017), 138--147.

\bibitem[KEL{\etalchar{*}}22]{unhealthyflowvolume}
\textsc{Kaszczewski P., Elwertowski M., Leszczyński J., Ostrowski T.,
  Kaszczewska J., Brzeziński T., Jarosz D., Świeczkowski Feiz S., Gałązka
  Z.}:
\newblock Volumetric flow assessment in extracranial arteries in patients with
  70-99\% internal carotid artery stenosis.
\newblock \emph{Diagnostics (Basel) 12}, 9 (Sep 2022).
\newblock \href {https://doi.org/10.3390/diagnostics12092216}
  {\path{doi:10.3390/diagnostics12092216}}.

\bibitem[KMM{\etalchar{*}}18]{kreiser2018survey}
\textsc{Kreiser J., Meuschke M., Mistelbauer G., Preim B., Ropinski T.}:
\newblock A survey of flattening-based medical visualization techniques.
\newblock In \emph{Computer Graphics Forum} (2018), vol.~37, pp.~597--624.
\newblock \href {https://doi.org/10.1111/cgf.13445}
  {\path{doi:10.1111/cgf.13445}}.

\bibitem[Kov15]{kovesi2015good}
\textsc{Kovesi P.}:
\newblock Good colour maps: How to design them, colorcet.com, 2015.
\newblock \href {https://doi.org/10.48550/ARXIV.1509.03700}
  {\path{doi:10.48550/ARXIV.1509.03700}}.

\bibitem[Kut01]{kutta1901beitrag}
\textsc{Kutta W.}:
\newblock {Beitrag zur n\"aherungsweisen Integration totaler
  Differentialgleichungen}.
\newblock \emph{Z. Math. Phys. 46} (1901), 435--453.

\bibitem[LABFL09]{lesage2009review}
\textsc{Lesage D., Angelini E.~D., Bloch I., Funka-Lea G.}:
\newblock A review of {3D} vessel lumen segmentation techniques: Models,
  features and extraction schemes.
\newblock \emph{Medical Image Analysis 13}, 6 (2009), 819--845.

\bibitem[Lee13]{lee2013general}
\textsc{Lee W.}:
\newblock General principles of carotid doppler ultrasonography.
\newblock \emph{Ultrasonography 33}, 1 (dec 2013), 11--17.
\newblock \href {https://doi.org/10.14366/usg.13018}
  {\path{doi:10.14366/usg.13018}}.

\bibitem[LGW{\etalchar{*}}19]{li2019hemodynamic}
\textsc{Li C.-H., Gao B.-L., Wang J.-W., Liu J.-F., Li H., Yang S.-T.}:
\newblock Hemodynamic factors affecting carotid sinus atherosclerotic stenosis.
\newblock \emph{World Neurosurgery 121} (jan 2019), e262--e276.
\newblock \href {https://doi.org/10.1016/j.wneu.2018.09.091}
  {\path{doi:10.1016/j.wneu.2018.09.091}}.

\bibitem[LHH{\etalchar{*}}19]{lee2019effect}
\textsc{Lee S.~H., Han K.-S., Hur N., Cho Y.~I., Jeong S.-K.}:
\newblock The effect of patient-specific non-newtonian blood viscosity on
  arterial hemodynamics predictions.
\newblock \emph{Journal of Mechanics in Medicine and Biology 19}, 08 (2019),
  1940054.
\newblock \href {https://doi.org/10.1142/S0219519419400542}
  {\path{doi:10.1142/S0219519419400542}}.

\bibitem[LK16]{liskowski2016segmenting}
\textsc{Liskowski P., Krawiec K.}:
\newblock Segmenting retinal blood vessels with deep neural networks.
\newblock \emph{{IEEE} Transactions on Medical Imaging 35}, 11 (nov 2016),
  2369--2380.
\newblock \href {https://doi.org/10.1109/tmi.2016.2546227}
  {\path{doi:10.1109/tmi.2016.2546227}}.

\bibitem[LPTL20]{LOPES2020110019}
\textsc{Lopes D., Puga H., Teixeira J., Lima R.}:
\newblock Blood flow simulations in patient-specific geometries of the carotid
  artery: A systematic review.
\newblock \emph{Journal of Biomechanics 111} (2020), 110019.
\newblock \href {https://doi.org/10.1016/j.jbiomech.2020.110019}
  {\path{doi:10.1016/j.jbiomech.2020.110019}}.

\bibitem[MFW{\etalchar{*}}20]{mendieta2020importance}
\textsc{Mendieta J.~B., Fontanarosa D., Wang J., Paritala P.~K., McGahan T.,
  Lloyd T., Li Z.}:
\newblock The importance of blood rheology in patient-specific computational
  fluid dynamics simulation of stenotic carotid arteries.
\newblock \emph{Biomechanics and Modeling in Mechanobiology 19}, 5 (2020),
  1477--1490.
\newblock \href {https://doi.org/10.1007/s10237-019-01282-7}
  {\path{doi:10.1007/s10237-019-01282-7}}.

\bibitem[MGB{\etalchar{*}}19]{meuschke2019visual}
\textsc{Meuschke M., Gunther T., Berg P., Wickenhofer R., Preim B., Lawonn K.}:
\newblock Visual analysis of aneurysm data using statistical graphics.
\newblock \emph{{IEEE} Transactions on Visualization and Computer Graphics 25},
  1 (jan 2019), 997--1007.
\newblock \href {https://doi.org/10.1109/tvcg.2018.2864509}
  {\path{doi:10.1109/tvcg.2018.2864509}}.

\bibitem[MKP{\etalchar{*}}16]{Meuschke_2016_EuroVis}
\textsc{Meuschke M., Köhler B., Preim U., Preim B., Lawonn K.}:
\newblock Semi-automatic vortex flow classification in {4D} {PC-MRI} data of
  the aorta.
\newblock \emph{Computer Graphics Forum 35}, 3 (2016), 351--360.
\newblock \href {https://doi.org/https://doi.org/10.1111/cgf.12911}
  {\path{doi:https://doi.org/10.1111/cgf.12911}}.

\bibitem[MLW00]{mead2000variability}
\textsc{Mead G.~E., Lewis S.~C., Wardlaw J.~M.}:
\newblock Variability in doppler ultrasound influences referral of patients for
  carotid surgery.
\newblock \emph{European Journal of Ultrasound 12}, 2 (dec 2000), 137--143.
\newblock \href {https://doi.org/10.1016/s0929-8266(00)00111-7}
  {\path{doi:10.1016/s0929-8266(00)00111-7}}.

\bibitem[MVB{\etalchar{*}}16]{meuschke2016combined}
\textsc{Meuschke M., Voss S., Beuing O., Preim B., Lawonn K.}:
\newblock Combined visualization of vessel deformation and hemodynamics in
  cerebral aneurysms.
\newblock \emph{IEEE Transactions on Visualization and Computer Graphics 23}, 1
  (2016), 761--770.
\newblock \href {https://doi.org/10.1109/TVCG.2016.2598795}
  {\path{doi:10.1109/TVCG.2016.2598795}}.

\bibitem[MVB{\etalchar{*}}17]{meuschke2017glyph}
\textsc{Meuschke M., Vo{\ss} S., Beuing O., Preim B., Lawonn K.}:
\newblock Glyph-based comparative stress tensor visualization in cerebral
  aneurysms.
\newblock In \emph{Computer Graphics Forum} (2017), vol.~36, pp.~99--108.
\newblock \href {https://doi.org/10.1111/cgf.13171}
  {\path{doi:10.1111/cgf.13171}}.

\bibitem[MVE{\etalchar{*}}22]{Meuschke2022}
\textsc{Meuschke M., Voß S., Eulzer P., Janiga G., Arens C., Wickenhöfer R.,
  Preim B., Lawonn K.}:
\newblock {COMFIS} -- comparative visualization of simulated medical flow data.
\newblock In \emph{Eurographics Workshop on Visual Computing for Biology and
  Medicine} (2022).

\bibitem[MVN06]{manniesing2006vessel}
\textsc{Manniesing R., Viergever M.~A., Niessen W.~J.}:
\newblock Vessel enhancing diffusion: A scale space representation of vessel
  structures.
\newblock \emph{Medical Image Analysis 10}, 6 (dec 2006), 815--825.
\newblock \href {https://doi.org/10.1016/j.media.2006.06.003}
  {\path{doi:10.1016/j.media.2006.06.003}}.

\bibitem[MWvdV{\etalchar{*}}16]{moeskops2016deep}
\textsc{Moeskops P., Wolterink J.~M., van~der Velden B. H.~M., Gilhuijs K.
  G.~A., Leiner T., Viergever M.~A., I{\v{s}}gum I.}:
\newblock Deep learning for multi-task medical image segmentation in multiple
  modalities.
\newblock In \emph{Medical Image Computing and Computer-Assisted Intervention
  {\textendash} {MICCAI} 2016}. Springer International Publishing, 2016,
  pp.~478--486.
\newblock \href {https://doi.org/10.1007/978-3-319-46723-8_55}
  {\path{doi:10.1007/978-3-319-46723-8_55}}.

\bibitem[NHC{\etalchar{*}}20]{nisco2020impact}
\textsc{Nisco G.~D., Hoogendoorn A., Chiastra C., Gallo D., Kok A.~M.,
  Morbiducci U., Wentzel J.~J.}:
\newblock The impact of helical flow on coronary atherosclerotic plaque
  development.
\newblock \emph{Atherosclerosis 300} (may 2020), 39--46.
\newblock \href {https://doi.org/10.1016/j.atherosclerosis.2020.01.027}
  {\path{doi:10.1016/j.atherosclerosis.2020.01.027}}.

\bibitem[OJMN{\etalchar{*}}19]{oeltze2019generation}
\textsc{Oeltze-Jafra S., Meuschke M., Neugebauer M., Saalfeld S., Lawonn K.,
  Janiga G., Hege H.-C., Zachow S., Preim B.}:
\newblock Generation and visual exploration of medical flow data: Survey,
  research trends and future challenges.
\newblock In \emph{Computer Graphics Forum} (2019), vol.~38, pp.~87--125.
\newblock \href {https://doi.org/10.1111/cgf.13394}
  {\path{doi:10.1111/cgf.13394}}.

\bibitem[OLK{\etalchar{*}}14]{oeltze2014blood}
\textsc{Oeltze S., Lehmann D.~J., Kuhn A., Janiga G., Theisel H., Preim B.}:
\newblock Blood flow clustering and applications in virtual stenting of
  intracranial aneurysms.
\newblock \emph{IEEE Transactions on Visualization and Computer Graphics 20}, 5
  (2014), 686--701.
\newblock \href {https://doi.org/10.1109/TVCG.2013.2297914}
  {\path{doi:10.1109/TVCG.2013.2297914}}.

\bibitem[Pan07]{panerai2007Cerebral}
\textsc{Panerai R.~B.}:
\newblock Cerebral autoregulation: From models to clinical applications.
\newblock \emph{Cardiovascular Engineering 8}, 1 (nov 2007), 42--59.
\newblock \href {https://doi.org/10.1007/s10558-007-9044-6}
  {\path{doi:10.1007/s10558-007-9044-6}}.

\bibitem[RGB{\etalchar{*}}19]{rajsic2019economic}
\textsc{Rajsic S., Gothe H., Borba H., Sroczynski G., Vujicic J., Toell T.,
  Siebert U.}:
\newblock Economic burden of stroke: a systematic review on post-stroke care.
\newblock \emph{The European Journal of Health Economics 20}, 1 (2019),
  107--134.
\newblock \href {https://doi.org/10.1007/s10198-018-0984-0}
  {\path{doi:10.1007/s10198-018-0984-0}}.

\bibitem[RiK21]{ryu2021efficient}
\textsc{Ryu J., ichiro Kamata S.}:
\newblock An efficient computational algorithm for hausdorff distance based on
  points-ruling-out and systematic random sampling.
\newblock \emph{Pattern Recognition 114} (jun 2021), 107857.
\newblock \href {https://doi.org/10.1016/j.patcog.2021.107857}
  {\path{doi:10.1016/j.patcog.2021.107857}}.

\bibitem[RPH{\etalchar{*}}24]{richter2023}
\textsc{Richter K., Probst T., Hundertmark A., Eulzer P., Lawonn K.}:
\newblock Longitudinal wall shear stress evaluation using centerline projection
  approach in the numerical simulations of the patient-based carotid artery.
\newblock \emph{Computer Methods in Biomechanics and Biomedical Engineering
  27}, 3 (2024), 347--364.
\newblock \href {https://doi.org/10.1080/10255842.2023.2185478}
  {\path{doi:10.1080/10255842.2023.2185478}}.

\bibitem[Run95]{runge1895numerische}
\textsc{Runge C.}:
\newblock {{\"U}ber die numerische Aufl{\"o}sung von Differentialgleichungen}.
\newblock \emph{Mathematische Annalen 46}, 2 (1895), 167--178.
\newblock \href {https://doi.org/10.1007/BF01446807}
  {\path{doi:10.1007/BF01446807}}.

\bibitem[SBdB16]{saha2016survey}
\textsc{Saha P.~K., Borgefors G., di~Baja G.~S.}:
\newblock A survey on skeletonization algorithms and their applications.
\newblock \emph{Pattern Recognition Letters 76} (jun 2016), 3--12.
\newblock \href {https://doi.org/10.1016/j.patrec.2015.04.006}
  {\path{doi:10.1016/j.patrec.2015.04.006}}.

\bibitem[SCA{\etalchar{*}}16]{sousa2016computational}
\textsc{Sousa L.~C., Castro C.~F., Ant{\'{o}}nio C.~C., Sousa F., Santos R.,
  Castro P., Azevedo E.}:
\newblock Computational simulation of carotid stenosis and flow dynamics based
  on patient ultrasound data {\textendash} a new tool for risk assessment and
  surgical planning.
\newblock \emph{Advances in Medical Sciences 61}, 1 (mar 2016), 32--39.
\newblock \href {https://doi.org/10.1016/j.advms.2015.07.009}
  {\path{doi:10.1016/j.advms.2015.07.009}}.

\bibitem[SHS18]{SZAJER201862}
\textsc{Szajer J., Ho-Shon K.}:
\newblock A comparison of {4D} flow {MRI}-derived wall shear stress with
  computational fluid dynamics methods for intracranial aneurysms and carotid
  bifurcations: A review.
\newblock \emph{Magnetic Resonance Imaging 48} (2018), 62--69.
\newblock \href {https://doi.org/10.1016/j.mri.2017.12.005}
  {\path{doi:10.1016/j.mri.2017.12.005}}.

\bibitem[SRC{\etalchar{*}}20]{sun2020zernet}
\textsc{Sun Z., Rooke E., Charton J., He Y., Lu J., Baek S.}:
\newblock {ZerNet}: Convolutional neural networks on arbitrary surfaces via
  zernike local tangent space estimation.
\newblock \emph{Computer Graphics Forum 39}, 6 (may 2020), 204--216.
\newblock \href {https://doi.org/10.1111/cgf.14012}
  {\path{doi:10.1111/cgf.14012}}.

\bibitem[TGK{\etalchar{*}}16]{tominski2016interactive}
\textsc{Tominski C., Gladisch S., Kister U., Dachselt R., Schumann H.}:
\newblock Interactive lenses for visualization: An extended survey.
\newblock \emph{Computer Graphics Forum 36}, 6 (may 2016), 173--200.
\newblock \href {https://doi.org/10.1111/cgf.12871}
  {\path{doi:10.1111/cgf.12871}}.

\bibitem[ULP{\etalchar{*}}07]{Ugray2007}
\textsc{Ugray Z., Lasdon L., Plummer J., Glover F., Kelly J., Mart\'{\i} R.}:
\newblock Scatter search and local nlp solvers: A multistart framework for
  global optimization.
\newblock \emph{INFORMS Journal on Computing 19}, 3 (2007), 328--340.
\newblock \href {https://doi.org/10.1287/ijoc.1060.0175}
  {\path{doi:10.1287/ijoc.1060.0175}}.

\bibitem[VCFJT06]{vignonclementel}
\textsc{Vignon-Clementel I.~E., Figueroa C.~A., Jansen K.~E., Taylor C.~A.}:
\newblock Outflow boundary conditions for three-dimensional finite element
  modeling of blood flow and pressure in arteries.
\newblock \emph{Computer Cethods In Applied Mechanics And Engineering 195},
  29-32 (June 2006), 3776--3796.

\bibitem[VPvP{\etalchar{*}}14]{vilanova2014visual}
\textsc{Vilanova A., Preim B., van Pelt R., Gasteiger R., Neugebauer M.,
  Wischgoll T.}:
\newblock Visual exploration of simulated and measured blood flow.
\newblock In \emph{Scientific Visualization}. Springer, London, 2014,
  pp.~305--324.
\newblock \href {https://doi.org/10.1007/978-1-4471-6497-5_25}
  {\path{doi:10.1007/978-1-4471-6497-5_25}}.

\bibitem[VRF{\etalchar{*}}15]{vitello2015blood}
\textsc{Vitello D.~J., Ripper R.~M., Fettiplace M.~R., Weinberg G.~L., Vitello
  J.~M.}:
\newblock Blood density is nearly equal to water density: A validation study of
  the gravimetric method of measuring intraoperative blood loss.
\newblock \emph{Journal of Veterinary Medicine 2015} (2015), 1--4.
\newblock \href {https://doi.org/10.1155/2015/152730}
  {\path{doi:10.1155/2015/152730}}.

\bibitem[Wei00]{healthyflowvolume}
\textsc{Weigand C.}:
\newblock \emph{Strömungsanalysen in der Karotisbifurkation}.
\newblock PhD thesis, Technische Universität München, 2000.

\bibitem[WXG{\etalchar{*}}16]{wu2016deep}
\textsc{Wu A., Xu Z., Gao M., Buty M., Mollura D.~J.}:
\newblock Deep vessel tracking: A generalized probabilistic approach via deep
  learning.
\newblock In \emph{2016 {IEEE} 13th International Symposium on Biomedical
  Imaging ({ISBI})} (apr 2016), {IEEE}.
\newblock \href {https://doi.org/10.1109/isbi.2016.7493520}
  {\path{doi:10.1109/isbi.2016.7493520}}.

\bibitem[ZTP{\etalchar{*}}21]{zhou2021fully}
\textsc{Zhou T., Tan T., Pan X., Tang H., Li J.}:
\newblock Fully automatic deep learning trained on limited data for carotid
  artery segmentation from large image volumes.
\newblock \emph{Quantitative Imaging in Medicine and Surgery 11}, 1 (jan 2021),
  67--83.
\newblock \href {https://doi.org/10.21037/qims-20-286}
  {\path{doi:10.21037/qims-20-286}}.

\end{thebibliography}


\end{document}